# Low Complexity Component Nonlinear Distortions Mitigation Scheme for Probabilistically Shaped 64-QAM Signals

Yiwen Wu[(1)], Mengfan Fu[(1)], Huazhi Lun[(1)], Lilin Yi[(1)], Weisheng Hu[(1)] and Qunbi Zhuge[*(1)]

[(1)] State Key Laboratory of Advanced Optical Communication Systems and Networks, Department of Electronic Engineering, Shanghai Jiao Tong University, Shanghai, 200240, China, qunbi.zhuge@sjtu.edu.cn

**Abstract** *We propose a degenerated hierarchical look-up table (DH-LUT) scheme to compensate component nonlinearities. For probabilistically shaped 64-QAM signals, it achieves up to 2-dB SNR improvement, while the size of table is only 8.59% compared to the conventional LUT method.*

## Introduction

Due to the rapid development of internet applications, the needed data throughput over intra- and inter-datacentre optical links with up to 10-km transmission distance is exponentially increasing[1]-[2]. To further improve the capacity of datacentre networks, higher order modulation formats are required such as 32-quadrature amplitude modulation (32-QAM) and 64-QAM. For these formats, the nonlinear distortions of optical and electrical components become one of the major impairments[3], since the fiber link impairment is insignificant in short-reach links. In the meantime, datacentre links cannot afford high power consumption for optical transceivers. Therefore, a low complexity scheme to mitigate component nonlinearities is urgently needed. Recently, look-up table (LUT) based nonlinear mitigation has attracted many attentions as an efficient method[4]-[6]. However, the size of table exponentially increases as the modulation order gets higher. To reduce the complexity of LUT, a modified LUT was proposed for 16-pulse amplitude modulation (16-PAM) which only compensates for high amplitude symbols[5]. Nevertheless, this method only halves the size of table, and further reduction is needed, especially for higher modulation format.

In this paper, we propose a degenerated hierarchical LUT (DH-LUT) scheme to compensate the component nonlinearities for probabilistically shaped 64-QAM (PS-64-QAM) signals. This scheme adopts a hierarchical structure and a table degeneration scheme to reduce the size of table. Experiments are conducted to investigate the performance of the DH-LUT, and up to 2-dB signal-to-noise ratio (SNR) improvement is demonstrated. Meanwhile, the required size of LUT is only 8.59% compared to the conventional full-size LUT method.

## Principle of DH-LUT

The proposed DH-LUT scheme is composed of two stages. First, a hierarchical LUT (H-LUT) is applied to split the large table into two small tables. Then, the table degeneration is performed to further reduce the size of the second LUT (LUT-2).

Fig. 1 depicts the block diagram of the H-LUT. The table is established based on the training method. The training signal $X_t(k)$ with a sufficient length is transmitted in the back-to-back (B2B) case. After the receiver side digital signal processing (DSP), the received signal $X'_t(k)$ is obtained. Then to process each symbol, we construct a *M*-symbol block extracted from the training signal $X_t(k)$. In the H-LUT scheme, the *M*-symbol block is split into two parts, which consist of the current symbol and the past (*M*-1) symbols. The current symbol is used as the input of the first look-up table (LUT-1), and the past (*M*-1) symbols are used as the input of the LUT-2.

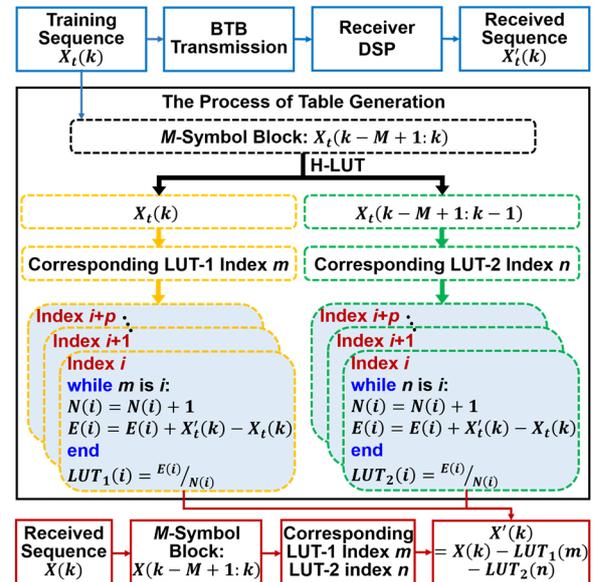

**Fig. 1:** The block diagram of the H-LUT scheme.

First, we group all the *M*-symbol blocks with the first symbol corresponding to a specific input of LUT-1. Then the errors of the first symbol of these blocks after transmission are averaged. As a result, the output of LUT-1 is obtained. A similar procedure is conducted to obtain the output of LUT-2 by grouping all the *M*-symbol blocks with the past (*M*-1) symbols corresponding to a specific input of LUT-2.

After the tables are established, during regular transmissions we extract the *M*-symbol blocks from the received signal and input the first symbol and past (*M*-1) symbols to LUT-1 and LUT-2, respectively. Then the distortions obtained by LUT-1 and LUT-2 are subtracted from the received signal to compensate the nonlinearities.

In this paper, we investigate the modulation format of 64-QAM. By setting *M* to 3, the sizes of LUT-1 and LUT-2 are 8 and 64, respectively. To further reduce the size of LUT-2, we propose the DH-LUT method to degenerate the LUT-2. We first degenerate the in-phase or quadrature (I/Q) signal of 64-QAM, i.e. the 8-amplitude shift keying (8-ASK) signal into less points. Then the reduced size table can be obtained by weighted averaging the original table's elements with the specific indexes. In Fig. 2(a) and (b), we plot the inputs of the original LUT-2 of H-LUT and the degenerated LUT-2 of 2-DH-LUT, respectively. As the figure shows, the degenerated table with a size of 4 can be realized by degenerating the 8-ASK symbols [-7, -5, -3, -1, 1, 3, 5, 7] into the 2-ASK symbols [-1, 1].

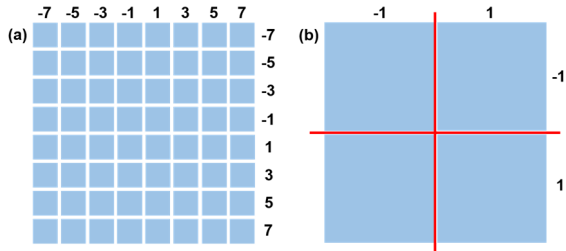

**Fig. 2:** (a) The original LUT-2 of H-LUT; (b) The degenerated LUT-2 of 2-DH-LUT.

To achieve better compensation performance, the degenerated schemes should be optimized. Because the component nonlinearities are more serious for symbols with high amplitudes, then a better scheme is to distribute more degenerated symbols at high amplitudes. On the contrary, since the PS signals have more symbols with low amplitudes, then distributing more degenerated symbols at low amplitudes is a better choice. Therefore, there exits an optimal degeneration scheme to balance the effects of component nonlinearities and probabilistic shaping. In order to find the optimal degeneration scheme, we define a performance parameter $\eta$ to evaluate the performance of DH-LUT by calculating the weighted summation of the Euclidean distance between the original and degenerated tables. The performance parameter $\eta$ is given as

$$\eta = \frac{\sum_{i \in j}|T2(i)-D[T2(j)]|^2*P(i)}{\sum_{i \in j}|T2(i)|^2*P(i)} \quad (1)$$

$$D[T(j)] = \frac{\sum_{i \in j}T2(i)*P(i)}{\sum_{i \in j}P(i)} \quad (2)$$

where $D[\cdot]$ is the degeneration function, $T2[i]$ and $P[i]$ are the $i$-th element of LUT-2 and its probability weight, respectively. As the 8-ASK signals can be degenerated into 2-ASK, 4-ASK or 6-ASK, the size of LUT-2 can be degenerated into 4, 16 or 36.

**Experimental setup**

The experimental setup is shown in Fig. 3. At the transmitter side, a single channel 20-GBaud dual polarization 64-QAM (DP-64-QAM) signal with root-raised cosine (RRC) pulse shaping (roll-off factor = 0.1) and probabilistic shaping is first generated. The spectrum efficiency (SE) is 5.8-bit/symbol. The transmitted waveforms are generated offline in MATLAB and then uploaded to an arbitrary waveform generator (AWG) with a sampling rate of 80-GSa/s. Afterwards, the four output electrical signals are driven by radio frequency (RF) drivers followed by a DP-I/Q modulator. The linewidth of the tunable laser is 100-kHz and the central frequency is 193.41-THz. Then the modulated signal is amplified by an Erbium-doped fiber amplifier (EDFA). After 10-km transmission over standard single mode fiber (SSMF), the received optical power (ROP) of the received signal is adjusted by a variable optical attenuator (VOA). Afterwards, a 4-channel real-time digital storage oscilloscope (DSO) with a sampling rate of 100-GSa/s is used to digitize the signal. In the receiver side DSP, the IQ errors are first compensated. Then the frequency offset compensation (FOC) is performed, followed by the decision-directed least mean square (DD-LMS) equalization and carrier phase recovery (CPR). Then the LUT-assisted compensation for each I/Q lane is performed. Note that the table of each I/Q lane is trained independently, and the component nonlinearities of each I/Q lane is also compensated independently. At last, the bit error ratio (BER) and SNR are calculated to evaluate the transmission performance.

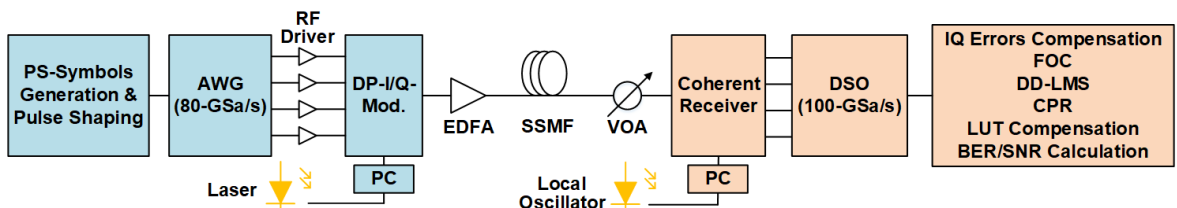

**Fig. 3:** Experimental setup. PC: polarization controller.

## Results and discussions

First, the performance of the conventional full-size LUT and our proposed H-LUT schemes is investigated to illustrate the impact of the hierarchical operation. Fig. 4 shows the BER versus ROP for the LUT and H-LUT. The experimental results with a ROP of -8-dBm show that the LUT and H-LUT decrease the BER from $1.94\times10^{-2}$ to $5.9\times10^{-3}$ and $6.4\times10^{-3}$, respectively. The performance of H-LUT is slightly worse than LUT, but the table size of H-LUT is only 14.06% relative to the latter.

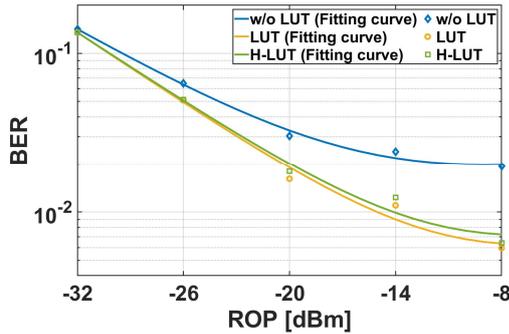

**Fig. 4:** BER vs. ROP for the LUT and H-LUT schemes.

Next, we calculate the values of $\eta$ for all possible degeneration schemes, and find the optimal degeneration scheme with the minimal value of $\eta$. As shown in Fig. 5, the performance metric of the optimal degeneration schemes for different table sizes can be obtained. The degenerated constellations and corresponding LUT-2 are also shown in this figure. Since a lower $\eta$ means a better compensation performance, we observe that the compensation gain increases as the table size gets larger.

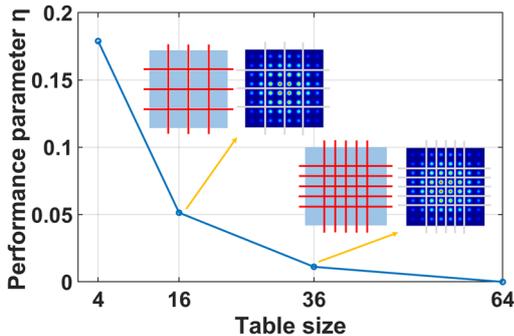

**Fig. 5:** Performance metrics with the optimal LUT degeneration for different table sizes.

Then we investigate the performance of the DH-LUT with various table sizes. In Fig. 6, we show the BER versus ROP for DH-LUT with different table sizes. With a ROP of -8-dBm, the 2-DH-LUT, 4-DH-LUT and 6-DH-LUT can decrease the BER to $7.8\times10^{-3}$, $6.7\times10^{-3}$ and $6.5\times10^{-3}$, respectively. The table size of 6-DH-LUT is only 8.59% of the LUT method, while its performance is comparable.

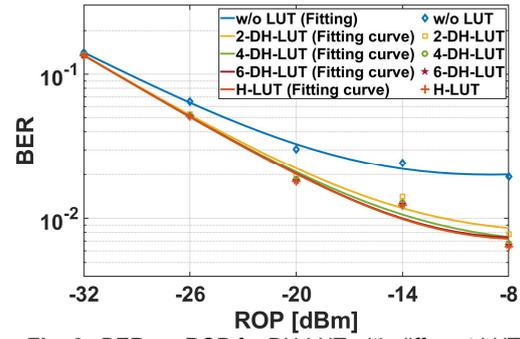

**Fig. 6:** BER vs. ROP for DH-LUT with different LUT-2.

At last, we investigate the SNR versus ROP for our schemes. As shown in Fig. 7, the SNR improvements of 6-DH-DLUT and LUT are 2-dB and 2.11-dB, respectively, compared to the system without compensation at a ROP of -8-dBm. The performance penalty of 6-DH-LUT is only 0.11-dB compared to the LUT method, while it significantly reduces the complexity.

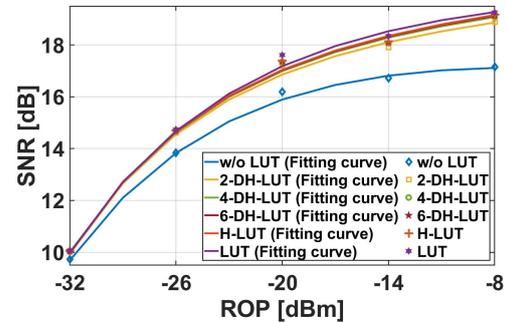

**Fig. 7:** SNR vs. ROP for DH-LUT with different LUT-2.

## Conclusion

In this paper, we propose a DH-LUT scheme to mitigate component nonlinearities. This method adopts a hierarchical structure and table degeneration to reduce the implementation complexity. In experiments with PS-64-QAM signals, the results at the ROP of -8-dBm show that the proposed 6-DH-LUT can achieve 2-dB improvement of SNR. It reduces the size of table to 8.59% relative to a full-size LUT method.

## Acknowledgements
This work was supported by National Key R&D Program of China (2018YFB1801200), NSFC (61801291), and Shanghai Rising-Star Program (19QA1404600).